\documentclass[prd,superscriptaddress,amsfonts,amssymb,amsmath,showpacs,twocolumn,nofootinbib]{revtex4-2}
\usepackage{bm}
\usepackage{amsfonts}
\usepackage{latexsym}
\usepackage{graphicx}
\usepackage{array}
\usepackage{amsmath,amssymb}
\usepackage{palatino}
\usepackage{xcolor} 
\usepackage{mathpazo}
\usepackage{xcolor}
\usepackage{rotating}
\usepackage{adjustbox}
\usepackage{tensor}
\usepackage{textcomp}
\usepackage{float}
\usepackage{booktabs}
\usepackage{dcolumn}
\usepackage[titletoc]{appendix}
\usepackage{booktabs}
\usepackage{multirow}
\usepackage{hyperref}
\hypersetup{colorlinks,citecolor=blue}
\usepackage{amsmath}
\usepackage{xcolor}
\usepackage{epsfig}
\usepackage{caption}
\usepackage{subcaption}
\usepackage{commath}
\usepackage[table]{xcolor}

\hypersetup{colorlinks,citecolor=magenta}
\hypersetup{colorlinks=true,linkcolor=magenta,filecolor=magenta,    urlcolor=blue}

\def\be{\begin{equation}}
\def\ee{\end{equation}}
\def\bea{\begin{eqnarray}}
\def\eea{\end{eqnarray}}

\definecolor{owngreen}{rgb}{0.0, 0.5, 0.0}

\begin{document}

\title{Testing the Direction of Dark Sector Interaction Using Late-Time Datasets}

\author{Anil Kandel}
\email{anil.kandel3227@gmail.com}
\affiliation{Department of Physics, St.Xavier’s College, Tribhuvan University, Kathmandu, Nepal}

\author{Phongpichit Channuie}
\email{phongpichit.ch@mail.wu.ac.th (Corresponding Author)}
\affiliation{School of Science \& College of Graduate Studies, Walailak University, Nakhon Si Thammarat, 80160, Thailand}

\author{Ertan G\"{u}dekli}
\email{gudekli@istanbul.edu.tr}
\affiliation{Department of Physics, Faculty of Science, Istanbul University, 34134, Istanbul, Turkey}

\author{Farruh Atamurotov}
\email{atamurotov@yahoo.com}
\affiliation{Kimyo International University in Tashkent, Shota Rustaveli str. 156, Tashkent 100121, Uzbekistan}

\begin{abstract}
In this work, we constrain the phenomenological interacting dark energy $\xi$IDE model using baryon acoustic oscillation measurements from DESI Data Release 1 and Data Release 2, combined with cosmic chronometer measurements, compressed CMB likelihoods, and three different Type Ia supernova compilations: Pantheon$^+$, DES-Dovekie, and Union3. We constrain the parameters of the $\xi$IDE model using a Markov Chain Monte Carlo analysis, using the nested sampling algorithm implemented in the \texttt{dynesty} package through the cosmological inference code \texttt{SimpleMC}. Compared to $\Lambda$CDM, the $\xi$IDE model favors slightly higher values of the Hubble constant, leading to a reduction in the Hubble tension. However, the tension is not completely resolved for any dataset combination. The parameter $\xi$ is consistently constrained to values close to its $\Lambda$CDM prediction, indicating no statistically significant deviation from the standard cosmological scenario. Similarly, except for the CMB + DESI DR1 + CC dataset combination, the dark energy equation-of-state parameter prefers values of $w_{\rm X}>-1$, showing a preference for quintessence-like behavior and providing evidence for dynamical dark energy. The interaction parameter combination, $\xi+3w_{\rm X}$, is found to be negative for all dataset combinations, corresponding to energy transfer from dark energy to dark matter. The strongest preference is obtained for the CMB + DESI DR2 + CC + Union3 dataset combination, yielding $\xi+3w_{\rm X}=-0.035\pm0.023$, which corresponds to a deviation of only $1.52\sigma$ from the non-interacting limit. Therefore, the current data provide only a weak indication of a non-zero dark-sector interaction. Model comparison based on the $N\sigma$ statistic shows weak evidence in favor of the interacting scenario, while Bayesian evidence consistently prefers the $\Lambda$CDM model. The current cosmological observations do not provide compelling evidence for an interaction in the dark sector, although all dataset combinations consistently favor a mild energy transfer from dark energy to dark matter.
\end{abstract}

\maketitle

%\tableofcontents
\section{Introduction}\label{sec_1} 
The discovery of accelerated expansion of the Universe was made in 1998 by two independent groups~\cite{SupernovaSearchTeam:1998fmf,SupernovaCosmologyProject:1998vns}, using Type Ia supernovae as cosmological probes. This result demonstrated that a cosmological model containing only ordinary matter is insufficient to account for the observed acceleration. Consequently, it points to the existence of an additional component with large negative pressure, known as dark energy \cite{Padmanabhan:2002ji, Copeland:2006wr}. The simplest candidate for dark energy is the vacuum energy density characterized by an equation of state $w_{\Lambda} = p_{\Lambda}/\rho_{\Lambda} \equiv -1$. The energy density of a cosmic component evolves as $\rho_i \propto a^{-3(1+w_i)}$ during the expansion of the Universe. Consequently, the matter density scales as $\rho_m \propto a^{-3}$, while the cosmological constant density $\rho_{\Lambda}$ remains constant. The Universe described by a cosmological constant ($\Lambda$) as the dark energy component is typically referred to as the $\Lambda$CDM model, with CDM denoting cold dark matter.

Despite its minimal nature and remarkable empirical success, the $\Lambda$CDM model remains beset by profound tensions that continue to motivate the exploration of alternative cosmological frameworks. Among these is the infamous cosmological constant problem \cite{Weinberg:1988cp}, wherein theoretical estimates of the quantum vacuum energy density exceed the observed value $\rho_{\Lambda}^{\text{obs}} \sim (10^{-3}~\text{eV})^4$ by more than 120 orders of magnitude \cite{Copeland:2006wr}. Closely related is the cosmic coincidence problem \cite{Zlatev:1998tr}, which questions why dark energy and matter densities are of the same order precisely in the present epoch.

A recent challenge to the standard cosmological paradigm has emerged from observational indications of a time-dependent dark energy component. In particular, analyses of baryon acoustic oscillation data from Dark Energy Spectroscopic Instrument (DESI)~2024, in combination with Type~Ia supernova measurements, provide evidence for an evolving dark energy density relative to a cosmological constant at a statistical significance of $2.5$-$3.9\sigma$~\cite{adame2025desi}. The inclusion of the  DESI Data Release~2 (DESI DR2) in a joint analysis with Type Ia supernovae constraints further strengthens this trend, pushing the significance beyond the $4\sigma$ level~\cite{DESI:2025zgx}. Later re-analyzes of the Dark Energy Survey (DES) 5-year sample of Type Ia supernovae (DES-SN5YR) dataset, particularly those using the updated DES-Dovekie sample, indicate a reduced statistical preference for dynamical dark energy, with significance estimates around $3.2 \sigma$ instead of the previously reported $4.2 \sigma$ \cite{DES:2025sig}. Motivated by these developments, a number of recent studies have explored the implications of DESI DR2 measurements and found evidence for dynamical dark energy \cite{DESI:2024kob,ChaudharyEPJc,ChaudharyPDU,SharmaJHEAp,ChaudharyJHEAp,ChaudharyAPJs,ChaudharyAaA,DESI:2024aqx,ChaudharyAPj,RitikaJHEAp,DESI:2025fii}.

In response to these theoretical and observational tensions, a variety of alternative dark energy models have been proposed. Among them are scalar field models with a dynamical equation of state, such as quintessence \cite{ratra1988cosmological}, phantom \cite{caldwell2002phantom}, quintom \cite{feng2006quintom}, and k-essence \cite{Armendariz-Picon:2000ulo,Chiba:2002mw}. In addition, interacting dark energy (IDE) models have attracted significant attention as a natural extension of this framework, in which a non-gravitational coupling within the dark sector allows energy exchange between dark matter and dark energy. Interacting dark energy (IDE) models, which introduce possible couplings between dark energy (DE) and dark matter (DM), have also been suggested and studied as a means to address the coincidence problem \cite{Amendola:1999er,Zimdahl:2001ar,Chimento:2003iea,Guo:2004xx,cao2011testing,Wei:2006ut}. Beyond these early studies, several interacting DE models with different forms of DM-DE coupling have also been proposed and extensively investigated in the literature \cite{benisty2024late,Giare:2024smz,Li:2025owk,Clemson:2011an,Li:2014eha,Xia:2016vnp,van2025linear,vanderWesthuizen:2025mnw,vanderWesthuizen:2025rip,silva2025new,zhai2025low,silva2024non,giare2024tightening,sabogal2024quantifying,forconi2024double,escamilla2023model,zhai2023consistent,KumarSharma2026}. These studies have reported a wide range of results, from constraints fully consistent with the non-interacting $\Lambda$CDM scenario to mild indications of energy transfer within the dark sector. However, the inferred interaction strength and direction remain strongly dependent on the adopted interaction model and observational datasets. Consequently, establishing the existence and direction of a dark-sector interaction remains an open question in modern cosmology.

Among the various interacting dark energy scenarios, a particularly simple and widely studied phenomenological model is the $\xi$IDE framework proposed by Dalal et al.~\cite{Dalal:2001dt}. In this model, the ratio between dark energy and dark matter evolves as $\rho_{\rm DE}/\rho_{\rm DM}\propto a^{\xi}$, where the parameter $\xi$ quantifies the severity of the cosmic coincidence problem. The standard $\Lambda$CDM cosmology is recovered for $\xi=3$ and $w=-1$, while deviations from these values can indicate the presence of an interaction within the dark sector. Owing to its simplicity and clear physical interpretation, the $\xi$IDE model provides a useful phenomenological framework for investigating the existence and direction of dark-sector interactions using cosmological observations.

Determining the direction of energy transfer within the dark sector is crucial, since it dictates whether the coincidence problem is alleviated or exacerbated. Despite the growing interest in interacting dark energy models, the direction and strength of dark-sector interactions remain poorly constrained. In particular, the impact of the latest DESI DR2 measurements on these quantities has not yet been thoroughly investigated within the $\xi$IDE framework. Given the improved statistical precision and the emerging evidence for dynamical dark energy from DESI observations, it is important to revisit this model using the latest cosmological datasets. In this work, we investigate the phenomenological $\xi$IDE model, placing particular emphasis on the implications of the DESI DR2 measurements. The transition from DESI DR1 to DR2 provides a valuable opportunity to examine the stability of dark-sector constraints and to assess how the improved observational precision and stronger preference for dynamical dark energy affect the inferred interaction strength and direction. By comparing the constraints obtained from the two DESI data releases, we investigate how the inferred interaction strength and direction evolve with the inclusion of the latest observational data. Such a comparison provides an important test of the robustness and cosmological viability of interacting dark-sector scenarios. 

Our paper is organized as follows. In Section~\ref{sec_2}, we present the cosmological framework of the $\xi$IDE model and the corresponding background evolution equations. In Section~\ref{sec_3}, we describe the observational datasets, methodology, and parameter estimation procedure, and present the resulting cosmological constraints. Finally, in Section~\ref{sec_4}, we discuss the implications of our results and summarize the main conclusions of this work.

\section{Theoretical Background}\label{sec_2}
\label{sect: model}
We consider interacting dark sector models in which DM and DE exchange energy. In these scenarios, the total energy-momentum tensor of the dark sector remains conserved, although the individual DM and DE components are no longer independently conserved. The corresponding conservation equations are given by
\begin{equation}\label{eq:Qcovariant}
\nabla^\mu T_{\mu\nu}^{(DM)} = Q_{\nu}, \qquad
\nabla^\mu T_{\mu\nu}^{(\mathrm{\rm X})} = -Q_{\nu},
\end{equation}
where $Q_{\nu}$ represents the interaction four-vector governing the energy-momentum transfer between DM and DE. Consequently, the total dark sector energy-momentum tensor satisfies
\begin{equation}
\nabla^\mu \left(T_{\mu\nu}^{(DM)} + T_{\mu\nu}^{(\mathrm{\rm X})}\right)=0.
\end{equation}

Under the assumption that momentum transfer is negligible in the comoving frame, the interaction four-vector is assumed to be aligned with the four-velocity $u_\nu$,
\begin{equation}
Q_\nu = Q\,u_\nu .
\end{equation}
For a homogeneous and isotropic FLRW background, this takes the form
\begin{equation}
Q_\nu = (Q,\mathbf{0}),
\end{equation}
where $Q(t)$ denotes the rate of energy exchange between the dark sector components.

The temporal component ($\nu=0$) of Eq.~(\ref{eq:Qcovariant}) then leads to the modified continuity equations
\begin{equation}\label{eq:continuity_c}
\dot{\rho}_{DM} + 3H\,\rho_{DM} = Q,
\end{equation}
\begin{equation}\label{eq:continuity_DE}
\dot{\rho}_{\mathrm{\rm X}} + 3H(1+w_{\mathrm{\rm X}})\,\rho_{\mathrm{\rm X}} = -Q,
\end{equation}
where positive values of $Q$ indicate energy transfer from dark energy to DM, whereas negative values correspond to energy flow in the opposite direction.

In the present work, we consider a phenomenological interacting dark energy scenario in which the ratio between the dark energy and matter energy densities evolves as a power law of the scale factor \cite{Dalal:2001dt}, namely
\begin{equation}
\rho_{\rm X} \propto \rho_{\rm m} a^{\xi},
\end{equation}
where $\xi$ is a constant that quantifies the severity of
the coincidence problem. Within this framework, $\xi = 3$ reproduces the standard $\Lambda$CDM cosmology, whereas $\xi = 0$ corresponds to a self-similar solution in which the coincidence problem is absent. Intermediate values, $0 < \xi \leq 3$, describe scenarios in which the coincidence problem is progressively alleviated \cite{Pavon:2004xk}. In this paper, we adopt this phenomenological parameterization as the basis of our interacting dark sector analysis.
Considering the phenomenological %$\xi$IDE
model in a flat FLRW universe with $\Omega_{\rm X0} + \Omega_{\rm m0} = 1$, we can derive the interaction term,
\begin{equation}
	Q = -H\rho_{\rm m}(\xi +3w_{\rm X})\Omega_{X},
\end{equation}
where $\Omega_{X} = \frac{1-\Omega_{\rm m0}}{(1-\Omega_{\rm m0})+ \Omega_{\rm m0}(1+z)^\xi}$, the case $\xi+3w_{\rm X}=0$ (which implies $Q = 0$) corresponds to the standard cosmology without any interaction between dark energy and matter. Conversely, $\xi+3w_{\rm X} \neq 0$ indicates the non-standard cosmology. Furthermore, when $\xi+3w_{\rm X} > 0$ ($Q < 0$), it suggests that energy is transferred from dark matter to DE, which tends to exacerbate the coincidence problem. Conversely, if $\xi+3w_{\rm X} < 0$ ($Q > 0$), it indicates that the energy is transferred from DE to DM, which could potentially  mitigate the coincidence problem. The $H(z)$ function of the phenomenological model can be expressed as \citep{cao2011testing}

\begin{equation}
\begin{split}
E(z)^2 &\equiv \frac{H^2}{H_0^2} \\
&= (1+z)^3
\left[
\Omega_{m0}
+
(1-\Omega_{m0})(1+z)^{-\xi}
\right]^{-3w_X/\xi}.
\end{split}
\end{equation}
%%%%%%%%%%%%%%%%%%%%%%%%%%%%%%%%%%%%%%%%%%%%%%%%%%%%%%%%
\section{Dataset and Methodology}\label{sec_3}
We constrain the cosmological parameters of the IDE scenario through a Bayesian statistical analysis performed with the \texttt{SimpleMC}\footnote{\protect\url{https://github.com/ja-vazquez/SimpleMC.git}} cosmological inference framework. The exploration of the parameter space is carried out using the dynamic nested sampling \cite{Higson:2018cwj,Ashton:2022grj} algorithm, implemented through the \texttt{dynesty}\footnote{\protect\url{https://github.com/joshspeagle/dynesty.git}} \cite{speagle2020dynesty} Python package integrated within \texttt{SimpleMC}. In our analysis, the number of live points is selected according to the conventional choice $50 \times \mathrm{ndim}$, where $\mathrm{ndim}$ represents the number of independent free parameters of the model. The generated Monte Carlo chains are then processed using the \texttt{GetDist}\footnote{\url{https://github.com/cmbant/getdist}} package \cite{lewis2025getdist}to derive the marginalized one-dimensional (1D) posterior distributions and the associated 2D confidence regions.

We also estimate the Bayesian evidence, $\ln \mathcal{Z}$, which is calculated by the \texttt{dynesty} package using the nested sampling algorithm, which serves as a statistical indicator of the compatibility between cosmological models and observational data. The relative performance of two models $\mathcal{M}_i$ and $\mathcal{M}_j$ is quantified via the Bayes factor $\mathcal{B}_{ij} = \mathcal{Z}_i / \mathcal{Z}_j$. We report the logarithmic Bayes factor defined as $\ln \mathcal{B}_{ij} = \ln \mathcal{Z}_{i} - \ln \mathcal{Z}_{j},$ where $\ln \mathcal{Z}_{i}$ and $\ln \mathcal{Z}_{j}$ are the Bayesian evidence for models $i$ and $j$, respectively. A higher value of $\ln \mathcal{Z}$ indicates a more strongly favored model. To interpret the strength of the evidence, we adopt the revised Jeffreys' scale  \cite{kass1995bayes, trotta2008bayes}. According to this criterion, the strength of evidence can be interpreted as follows:
$\ln \mathcal{B}_{ij}< 1$ indicates inconclusive evidence; $1 \leq \ln \mathcal{B}_{ij} < 2.5$ suggests weak evidence;
$2.5 \leq \ln \mathcal{B}_{ij} < 5$ corresponds to moderate evidence; $5 \leq \ln \mathcal{B}_{ij} < 10$ signifies strong evidence; and $\ln\mathcal{B}_{ij} \geq 10$ denotes decisive evidence. Here, $i$ represents the $\Lambda$CDM model and $j$ represents the $\xi$IDE model.

In our analysis, we employ a multi-probe observational dataset for cosmological inference, incorporating the BAO measurements from the DESI DR1/DR2, together with  Cosmic Chronometers, Type Ia supernova observations and the compressed CMB likelihood. In the following, we provide more details about these datasets.

\begin{itemize}
\item \textbf{Cosmic Chronometers:} We use Cosmic Chronometer (CC) observations, which are based on massive and passively evolving galaxies characterized by old stellar populations and minimal ongoing star formation. These enable direct determinations of the Hubble expansion rate, $H(z)$, through the differential age method \cite{Jimenez:2001gg}. For our analysis, we use 15 measurements of the Hubble parameter, $H(z)$, covering the redshift range $0.1791 \leq z \leq 1.965$ \cite{moresco2012improved,Moresco:2015cya,Moresco:2016mzx} as these include the full covariance matrix accounting for both statistical and systematic uncertainties \cite{Moresco:2018xdr,Moresco:2020fbm}.

\item \textbf{Baryon Acoustic Oscillation:} {We use Baryon Acoustic Oscillation (BAO) measurements from DESI DR1 and DESI DR2 \cite{adame2025desi,DESI:2025zgx}. It combines data from multiple cosmological  tracers: Bright Galaxy Sample (BGS), Luminous Red Galaxies (LRG1–3), Emission Line Galaxies (ELG1–2), Quasars (QSO), and Lyman-$\alpha$ forests. The BAO constraints are provided in terms of the following dimensionless ratios: $D_M/r_d$, $D_H/r_d$, $D_V/r_d$, and $D_M/D_H$. Here, $D_H(z) = c/H(z)$ is the Hubble distance, $ D_M(z) = c \int_0^z \frac{dz'}{H(z')} $ is the comoving angular diameter distance, $D_V(z) \equiv [zD^2_M(z)D_H(z)]^{1/3}$ is the volume-averaged distance, and $r_d$ is the standard ruler length, the sound horizon at the drag epoch.}

\item \textbf{Type Ia supernovae:} We use the Pantheon$+$ sample \cite{brout2022pantheon+}, which consists of 1,701 light curves from 1,550 supernovae, spanning the redshift interval $0.001 \leq z \leq 2.26$.  Supernovae with redshift $z < 0.01$ are excluded, as observations in this regime are significantly affected by systematic uncertainties associated with peculiar velocities.  We use the DES-Dovekie dataset  \cite{DES:2025sig}, corresponding to the newly recalibrated  DES~SN5Y dataset \cite{abbott2024dark}, which contains a total of 1820 SNe~Ia. We also use the Union3 compilation , which includes 2087 cosmologically useful Type Ia supernovae drawn from 24 independent datasets and spanning the redshift range $0.050 \leq z \leq 2.26$ \cite{Rubin:2023jdq}. 

\item \textbf{Cosmic Microwave Background:} To incorporate CMB constraints, we adopt the compressed \textit{CamSpec} likelihood, modeled as a $3 \times 3$ Gaussian in the parameter space $\{\theta_*, \omega_b, \omega_{bc}\}$, where $\theta_*$ is the angular scale subtended by the sound horizon at recombination, and $\omega_b$ and $\omega_{bc}$ denote the physical energy densities of baryons and total cold matter, respectively. Since angular scale $\theta_* \propto r_s(z_*) / D_M(z_*)$, fixing $r_s(z_*)$ under the assumption of a $\Lambda$CDM cosmology prior to recombination enables $\theta_*$ to constrain the comoving angular diameter distance $D_M(z_*)$, whereas $\omega_{bc}$ constrains the matter density. A complete description of this likelihood construction is provided in Appendix A of \cite{DESI:2025zgx}.

\end{itemize}
%%%%%%%%%%%%%%%%%%%%%%%%%%%%%%%%%%%%%%%%%%%%%%%%%%%%%%

\begin{table}
\centering
\begin{tabular}{lll}
\hline
\textbf{Model} & \textbf{Parameter} & \textbf{Prior} \\
\hline
\multirow{3}{*}{\(\Lambda\)CDM} 
& \( \Omega_{m0} \) & \( \mathcal{U}[0.1, 0.5] \) \\
& \( \Omega_{b}h^2 \) & \( \mathcal{U}[0.02, 0.025] \) \\
& \( h = H_0/100 \) & \( \mathcal{U}[0.4, 0.9] \) \\
\hline
\multirow{2}{*}{$\xi$IDE} 
& \( w_{\rm X} \) & \( \mathcal{U}[-2.0, 2.0] \) \\
& \( \xi \) & \( \mathcal{U}[0.0, 4.0] \) \\
\hline
\end{tabular}
\caption{The parameters and their priors including uniform priors $\mathcal{U}$ and the reduced Hubble constant $h \equiv H_0/100$.}\label{tab_2}
\end{table}

%%%%%%%%%%%%%%%%%%%%%%%%%%%%%%%%%%%%%%%%%%%%%%%%%%%%%%%%%
\begin{figure*}
\begin{subfigure}{.4\textwidth}
\includegraphics[width=\linewidth]
{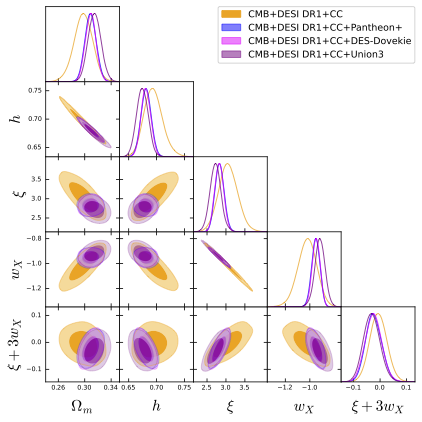}% width
\end{subfigure}
\hfil
\begin{subfigure}{.4\textwidth}
\includegraphics[width=\linewidth]
{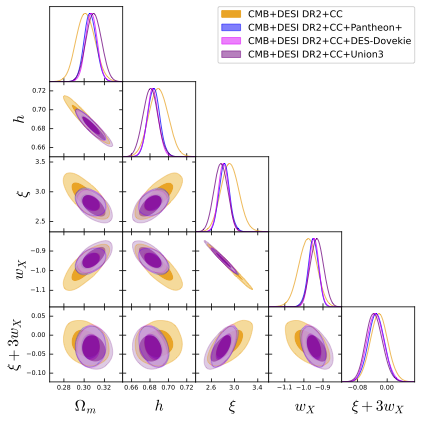}
% width
\end{subfigure}
\caption{This figure shows the corner plots of the $\xi$IDE model at the 68\% ($1\sigma$) and 95\% ($2\sigma$) confidence levels, obtained using DESI DR1 measurements (left panel) and DESI DR2 measurements (right panel) in combination with the compressed CMB likelihood, cosmic chronometer data, and different SNe~Ia compilations (Pantheon$+$, DES-Dovekie, and Union3).}\label{fig_1}
\end{figure*}
%%%%%%%%%%%%%%%%%%%%%%%%%%%%%%%%%%%%%%%%%%%%%%%%%%%%%%%%%
\begin{table*}
\setlength{\tabcolsep}{2.0pt}
\begin{tabular}{lcccccccc}
\hline
\textbf{Dataset / Model} & $h$ & $\Omega_m$ & $\xi$ & $w_{\rm X}$ & $\xi + 3 w_{\rm X}$ & $\ln \mathcal{B}_{i,j}$ & $\Delta{\chi^{2}_{\text{MAP}}}$ & $N\sigma$\\
\hline
\textbf{$\Lambda$CDM} \\
CMB + DESI DR1 + CC & $0.676 \pm 0.005$ & $0.310 \pm 0.007$ & --- & --- & $0$  & $0$ & $0$ & 0\\
CMB + DESI DR1  + CC + Pantheon$+$ & $0.673 \pm 0.005$ & $0.313 \pm 0.007$ & --- & --- & $0$ &   $0$  & $0$ & 0\\
CMB + DESI DR1 + CC + DES-Dovekie & $0.673 \pm 0.005$ & $0.314 \pm 0.006$ & --- & --- & $0$ & $0$ & $0$ & 0\\
CMB + DESI DR1 + CC + Union3  & $0.673 \pm 0.005$ & $0.314 \pm 0.007$ & --- & --- & $0$ & $0$ & $0$ &0\\
\addlinespace[4pt]
CMB + DESI DR2 + CC & $0.681 \pm 0.005$ & $0.303 \pm 0.006$ & --- & --- & $0$ & $0$ & $0$ & 0\\
CMB + DESI DR2  + CC + Pantheon$+$ & $0.679 \pm 0.004$ & $0.306 \pm 0.006$ & --- & --- & $0$ &  $0$ & $0$ & 0\\
CMB + DESI DR2 + CC + DES-Dovekie & $0.679 \pm 0.004$ & $0.307 \pm 0.005$ & --- & --- & $0$ & $0$ & $0$ & 0\\
CMB + DESI DR2 + CC + Union3  & $0.679 \pm 0.004$ & $0.306 \pm 0.005$ & --- & --- & $0$ & $0$ & $0$ &0\\
\hline
\textbf{$\xi$IDE} \\
CMB + DESI DR1 + CC & $0.694 \pm 0.016$ & $0.297 \pm 0.014$ & $3.05 \pm 0.25 $& $-1.021 \pm 0.076$ &$-0.006 \pm 0.032  $ &$5.38$ &$-1.21$ &$1.10$ \\
CMB + DESI DR1  + CC + Pantheon$+$ & $0.681 \pm 0.009$ & $0.308 \pm 0.008$ & $2.83 \pm 0.12$ & $-0.952 \pm 0.033$& $-0.024 \pm 0.030$ &$5.35$ & $-1.53$ & $1.24$\\
CMB + DESI DR1 + CC + DES-Dovekie & $0.680 \pm 0.008$ & $0.309 \pm 0.008$ & $2.82 \pm 0.11$ & $-0.948 \pm 0.032$ &$-0.025 \pm 0.031 $ & $4.38$ & $-3.60$ & $1.90$\\
CMB + DESI DR1 + CC + Union3  & $0.674 \pm 0.010$ & $0.314 \pm 0.009$ & $2.72 \pm 0.14$ & $-0.918 \pm 0.042$ &$ -0.031 \pm 0.031$ & $4.18$ & $-2.19$ &$1.48$\\
\addlinespace[4pt]
CMB + DESI DR2 + CC &$ 0.689 \pm 0.010 $  & $0.301 \pm 0.009$ & $2.92 \pm 0.16$ &$ -0.980 \pm 0.047$  &$ -0.022 \pm 0.024$ & $3.10$ & $-2.83$  & $ 1.68$\\
CMB + DESI DR2 + CC + Pantheon$+$  &$0.684 \pm 0.006 $  & $0.306 \pm 0.006$  &$ 2.82 \pm 0.09$ &$ -0.951 \pm 0.027$ & $-0.029 \pm 0.022 $ &$4.16$ & $-3.77$  & $1.94$  \\
CMB + DESI DR2 + CC + DES-Dovekie  &$0.683 \pm 0.006$  &$0.307 \pm 0.005 $  &$2.81 \pm 0.08   $ &$ -0.947 \pm 0.025$  &$-0.031 \pm 0.023 $ &$3.96 $ & $-4.14$ &$2.03$\\
CMB + DESI DR2 + CC + Union3  &$0.680 \pm 0.008 $  &     $ 0.309 \pm 0.007$ &$2.76 \pm 0.11$& $-0.932 \pm 0.034 $  &$-0.035 \pm 0.023 $ & $2.60 $ & $-4.26 $ & $2.06$\\
\hline
\end{tabular}
\caption{This table shows the mean values and corresponding errors at the 68\% (1$\sigma$) confidence level for the $\Lambda$CDM, and the $\xi$IDE models, obtained using DESI DR1/DR2 measurements in combination with the compressed CMB likelihood, cosmic chronometer data, and different SNe~Ia compilations (Pantheon$+$, DES-Dovekie, and Union3)}
\label{tab_2}
\end{table*}

\section{Results and Final Remarks}\label{sec_4}
In this section, we present the observational constraints and model comparison results for the $\xi$IDE scenario obtained using various combinations of CMB, DESI DR1/DR2, CC, Pantheon$^+$, DES-Dovekie, and Union3 datasets. We discuss the implications of the inferred cosmological parameters, examine the evidence for a possible interaction within the dark sector, assess the impact of the model on the Hubble tension, and compare its overall performance with that of the standard $\Lambda$CDM cosmology. In Fig.~\ref{fig_1}, we show the contour plots for the $\xi$IDE model obtained from the dataset combinations CMB + DESI + CC, CMB + DESI + CC + Pantheon$^{+}$, CMB + DESI + CC + DES-Dovekie, and CMB + DESI + CC + Union3. The contours derived using the DESI DR1 measurements are shown in the left column, while those obtained using the DESI DR2 measurements are shown in the right column. The diagonal panels show the marginalized one-dimensional (1D) posterior distributions of the model parameters, whereas the off-diagonal panels illustrate the corresponding 2D confidence contours. The inner and outer contours represent the 68\% ($1\sigma$) and 95\% ($2\sigma$) confidence regions, respectively.

From Table~\ref{tab_2}, which shows the mean values and corresponding $1\sigma$ uncertainties of the $\xi$IDE and $\Lambda$CDM models parameters, we first compare the constraints on the dimensionless Hubble parameter, $h$, obtained within the $\Lambda$CDM and $\xi$IDE frameworks. For the DESI DR1 dataset combinations, the $\xi$IDE model generally favors slightly higher values of $h$ than the corresponding $\Lambda$CDM scenario. For instance, the CMB + DESI DR1 + CC combination yields $h=0.694\pm0.016$ in the $\xi$IDE model, compared to $h=0.676\pm0.005$ in $\Lambda$CDM, corresponding to a mild deviation of $1.07\sigma$ between the two models. For the other DESI DR1 combinations, the differences remain below the $1\sigma$ level, with deviations of $0.78\sigma$, $0.74\sigma$, and $0.09\sigma$ for CMB + DESI DR1 + CC + Pantheon$^{+}$, CMB + DESI DR1 + CC + DES-Dovekie, and CMB + DESI DR1 + CC + Union3, respectively.

A similar trend is observed for the DESI DR2 dataset combinations, where the $\xi$IDE model continues to prefer mildly larger values of $h$. The largest value is obtained for the CMB + DESI DR2 + CC dataset combination, giving $h=0.689\pm0.010$ in $\xi$IDE, compared to $h=0.681\pm0.005$ in the corresponding $\Lambda$CDM case, with a deviation of $0.72\sigma$. For the remaining DESI DR2 combinations, the deviations between $\Lambda$CDM and $\xi$IDE are $0.69\sigma$, $0.55\sigma$, and $0.11\sigma$, respectively. Therefore, although the $\xi$IDE model tends to shift $h$ toward slightly larger values compared to $\Lambda$CDM, the differences between the two models remain statistically insignificant for all dataset combinations.

However, these values remain below the local distance-ladder measurement, $h^{\rm R22}=0.7304\pm0.0104$~\cite{Riess:2021jrx}. Within the $\Lambda$CDM framework, all dataset combinations shows a significant discrepancy with the R22 measurement, corresponding to tensions of approximately $4.3\sigma$--$5.0\sigma$. Specifically, the tensions are $4.71\sigma$, $4.97\sigma$, $4.97\sigma$, and $4.97\sigma$ for the DESI DR1 dataset combinations, and $4.28\sigma$, $4.61\sigma$, $4.61\sigma$, and $4.61\sigma$ for the corresponding DESI DR2 combinations.

In contrast, the $\xi$IDE model considerably reduces the tension with the local determination of the Hubble constant. For the DESI DR1 dataset combinations, the tensions decrease to $1.91\sigma$, $3.59\sigma$, $3.84\sigma$, and $3.91\sigma$, respectively, while for the DESI DR2 combinations they become $2.87\sigma$, $3.86\sigma$, $3.95\sigma$, and $3.84\sigma$. The most notable improvement is obtained for the CMB + DESI DR1 + CC dataset combination, where the tension is reduced from $4.71\sigma$ in $\Lambda$CDM to only $1.91\sigma$ in the $\xi$IDE scenario. Interestingly, replacing DESI DR1 with the DESI DR2 measurements increases the corresponding tension from $1.91\sigma$ to $2.87\sigma$, indicating that DESI DR2 weakens the preference for higher values of $h$ and consequently reduces the ability of the $\xi$IDE model to address the Hubble tension. Therefore, while the dark-sector interaction shifts the inferred value of $h$ closer to the local R22 measurement and significantly reduces the discrepancy relative to $\Lambda$CDM, it does not completely resolve the tension for any of the dataset combinations considered in this work.

We next compare the constraints on the present matter density parameter, $\Omega_m$, obtained within the $\Lambda$CDM and $\xi$IDE frameworks. In contrast to the Hubble parameter, the introduction of the dark-sector interaction has only a minor impact on the inferred values of $\Omega_m$. For the DESI DR1 dataset combinations, the $\xi$IDE model generally prefers slightly lower values of $\Omega_m$ than $\Lambda$CDM, with the largest difference found for the CMB + DESI DR1 + CC dataset combination, where $\Omega_m = 0.297 \pm 0.014$ in $\xi$IDE and $\Omega_m = 0.310 \pm 0.007$ in $\Lambda$CDM, corresponding to a deviation of only $0.83\sigma$. The remaining DESI DR1 combinations show even smaller differences, ranging from $0.00\sigma$ to $0.50\sigma$. A similar behavior is observed for the DESI DR2 dataset combinations, where the deviations between the two models remain below $0.35\sigma$. Overall, the constraints on $\Omega_m$ obtained in the $\Lambda$CDM and $\xi$IDE scenarios are highly consistent with one another, indicating that the dark-sector interaction does not significantly affect the determination of the present matter density from the current cosmological data.

We now discuss the model parameters of the $\xi$IDE model, beginning with the parameter $\xi$. This parameter describes how the ratio of dark energy to matter evolves with the expansion of the Universe, according to $\rho_{\rm X}/\rho_m \propto a^{\xi}$. As a result, $\xi$ provides a measure of the severity of the cosmic coincidence problem. In the standard $\Lambda$CDM cosmology, $\xi=3$, while $\xi=0$ corresponds to a self-similar solution in which the ratio between dark energy and matter remains constant over cosmic time. Therefore, values of $\xi$ smaller than 3 indicate a weaker coincidence problem than in the $\Lambda$CDM model, whereas values closer to 3 recover the standard cosmological behavior.

From Table~\ref{tab_2}, the $\xi$IDE model generally prefers values of $\xi$ slightly below the $\Lambda$CDM prediction. For the CMB + DESI DR1 + CC dataset, we obtain $\xi = 3.05 \pm 0.25$, which is fully consistent with the standard model expectation. However, the inclusion of SNe Ia data shifts the preferred value toward lower values, yielding $\xi = 2.83 \pm 0.12$, $\xi = 2.82 \pm 0.11$, and $\xi = 2.72 \pm 0.14$ for the Pantheon$^+$, DES-Dovekie, and Union3 compilations, respectively. A similar behavior is observed for the DESI DR2 analyses, where $\xi$ is constrained to be $2.92 \pm 0.16$, $2.82 \pm 0.09$, $2.81 \pm 0.08$, and $2.76 \pm 0.11$ for the CC-only, Pantheon$^+$, DES-Dovekie, and Union3 dataset combinations, respectively.

The obtained constraints favor values of $\xi$ in the range $2.7 \lesssim \xi \lesssim 2.9$, indicating a mild reduction of the coincidence problem compared to the standard $\Lambda$CDM scenario. However, the inferred values remain close to the canonical $\Lambda$CDM prediction of $\xi=3$, and therefore do not provide statistically significant evidence for a departure from the standard cosmological model.

Next, we consider the dark energy equation-of-state parameter, $w_{\rm X}$, which characterizes the dynamical properties of the dark energy component through the relation $w_{\rm X}=p_{\rm X}/\rho_{\rm X}$. In the standard $\Lambda$CDM model, dark energy is described by a cosmological constant with $w_{\rm X}=-1$. Values of $w_{\rm X}>-1$ correspond to a quintessence-like dark energy scenario, while $w_{\rm X}<-1$ indicates phantom behavior. Therefore, precise constraints on $w_{\rm X}$ provide important information about the nature of dark energy and its possible deviation from the cosmological constant paradigm.

The constraints on $w_{\rm X}$ are close to the cosmological constant value. For the CMB + DESI DR1 + CC dataset, we obtain $w_{\rm X}=-1.021\pm0.076$, which is fully consistent with $\Lambda$CDM. The addition of SNe Ia data shifts the preferred value toward the quintessence regime, yielding $w_{\rm X}=-0.952\pm0.033$, $w_{\rm X}=-0.948\pm0.032$, and $w_{\rm X}=-0.918\pm0.042$ for the Pantheon$^+$, DES-Dovekie, and Union3 compilations, respectively. A similar trend is observed for the DESI DR2 dataset combinations, where the constraints become $w_{\rm X}=-0.980\pm0.047$, $w_{\rm X}=-0.951\pm0.027$, $w_{\rm X}=-0.947\pm0.025$, and $w_{\rm X}=-0.932\pm0.034$ for the CC-only, Pantheon$^+$, DES-Dovekie, and Union3 analyses, respectively.

The constraints on $w_{\rm X}$ are confined to the range $-0.98 \lesssim w_{\rm X} \lesssim -0.92$, indicating the preference for a quintessence-like dark energy component. Nevertheless, the cosmological constant value, $w_{\rm X}=-1$, remains consistent with all dataset combinations within the $1$-$2\sigma$ confidence level. Consequently, the current observational data do not provide statistically significant evidence for a deviation from the $\Lambda$CDM paradigm, although they show a slight tendency toward a dynamical dark energy scenario with $w_{\rm X}>-1$.

We now turn to the combination $\xi + 3w_{\rm X}$, which plays a central role in the $\xi$IDE model since it directly determines the interaction term between dark matter and dark energy, $Q=-H\rho_m(\xi+3w_{\rm X})\Omega_{\rm X}.$ The standard $\Lambda$CDM cosmology corresponds to $\xi + 3w_{\rm X}=0$, implying the absence of any interaction within the dark sector. A negative value of $\xi + 3w_{\rm X}$ corresponds to $Q>0$, indicating an energy transfer from dark energy to dark matter, while a positive value implies the opposite energy flow. Therefore, constraining this parameter provides a direct test of the existence and direction of a possible dark-sector interaction.

The constraints on $\xi + 3w_{\rm X}$ consistently favor negative values across all dataset combinations. For the CMB + DESI DR1 + CC dataset, we obtain $\xi + 3w_{\rm X}=-0.006\pm0.032$, which is fully consistent with the non-interacting $\Lambda$CDM limit. The addition of SNe Ia data shifts the preferred value toward more negative values, yielding $\xi + 3w_{\rm X}=-0.024\pm0.030$, $-0.025\pm0.031$, and $-0.031\pm0.031$ for the Pantheon$^+$, DES-Dovekie, and Union3 compilations, respectively. A similar trend is observed for the DESI DR2 dataset combinations, where the constraints become $\xi + 3w_{\rm X}=-0.022\pm0.024$, $-0.029\pm0.022$, $-0.031\pm0.023$, and $-0.035\pm0.023$ for the CMB + DESI DR2 + CC, CMB + DESI DR2 + CC + Pantheon$^+$, CMB + DESI DR2 + CC + DES-Dovekie, and CMB + DESI DR2 + CC + Union3 dataset combinations, respectively.

The strongest indication of an interaction is obtained for the CMB + DESI DR2 + CC + Union3 dataset combination, which yields $\xi + 3w_{\rm X}=-0.035\pm0.023$, corresponding to a $1.52\sigma$ deviation from the non-interacting limit. Similarly, the CMB + DESI DR2 + CC + DES-Dovekie and CMB + DESI DR2 + CC + Pantheon$^+$ combinations favor negative values at the $1.35\sigma$ and $1.32\sigma$ levels, respectively. Although all dataset combinations consistently prefer $\xi + 3w_{\rm X}<0$, the statistical significance remains below the conventional detection threshold. Consequently, the current data provide only a weak indication of an energy transfer from dark energy to dark matter and do not yet constitute compelling evidence for a non-zero dark-sector interaction.

It is also interesting to compare our findings with previous studies. Early analyses based on $H(z)$ + BAO + CMB \citep{Cao:2010fb}, SNe Ia + BAO + CMB \citep{Chen:2010dk}, and \citep{Guo:2007zk} reported values of $\xi + 3w_{\rm X}$ consistent with zero but with substantially larger uncertainties. More recent analyses have yielded mixed results, ranging from $\xi + 3w_{\rm X}=0.03^{+1.35}_{-1.33}$ \citep{Zhu:2025lrk} and $-0.67^{+1.79}_{-1.58}$ \citep{Zheng:2022gfi}, which are fully consistent with the non-interacting scenario, to significantly negative values such as $-2.52\pm0.87$ \citep{Nong:2024bkr}, $-2.98\pm0.08$ \citep{Lebedev:2021tas}, $-3.10\pm0.40$ \citep{Wang:2016lxa}$,$ and $-3.54^{+0.12}_{-0.13}$ \citep{Salvatelli:2014zta}, which favor a substantial dark-sector interaction. Compared with these studies, our constraints are much closer to the $\Lambda$CDM limit and indicate only a weak preference for a negative interaction parameter. Nevertheless, the consistently negative central values across all dataset combinations provide an intriguing hint that the energy transfer, if present, is likely to proceed from dark energy to dark matter.

To quantify the preference for the $\xi$IDE model relative to $\Lambda$CDM, we consider the difference in the maximum-a-posteriori (MAP) goodness-of-fit statistic, denoted by $\Delta\chi^2_{\rm MAP}$. Assuming one additional degree of freedom, $\Delta\chi^2_{\rm MAP}$ is expected to follow a $\chi^2$ distribution with one degree of freedom. The corresponding Gaussian significance, $N\sigma$, is then obtained by converting the cumulative probability of the $\chi^2$ distribution into its equivalent normal-distribution significance through $\mathrm{CDF}_{\chi^2}\left(\Delta \chi^2_{\mathrm{MAP}} \, | \, 1~\mathrm{dof}\right)
= \frac{1}{\sqrt{2\pi}} \int_{-N}^{N} e^{-t^2/2} \, dt.$ Throughout this work, we adopt the following interpretation scale for the resulting significance: $N\sigma<1$ indicates inconclusive evidence, $1\leq N\sigma<2.5$ corresponds to weak evidence, $2.5\leq N\sigma<5$ indicates moderate evidence, $5\leq N\sigma<10$ corresponds to strong evidence, and $N\sigma\geq10$ is interpreted as decisive evidence in favor of the extended model.

For the DESI DR1 dataset combinations, the significance ranges from $1.10\sigma$ to $1.90\sigma$, while for the DESI DR2 combinations it ranges from $1.68\sigma$ to $2.06\sigma$. According to the adopted classification scheme, all of these values fall within the weak-evidence category ($1 \leq N\sigma < 2.5$). Interestingly, the DESI DR2 combinations generally yield slightly larger significance values than their DESI DR1 counterparts, with the highest preference obtained for the CMB + DESI DR2 + CC + Union3 dataset combination, corresponding to $N\sigma = 2.06$. Nevertheless, none of the dataset combinations reaches the moderate-evidence threshold of $2.5\sigma$. Therefore, although the $\xi$IDE model provides a better fit to the data than $\Lambda$CDM, the current observations indicate only weak statistical support for the interacting dark-sector scenario.

The Bayesian evidence values reported in Table~\ref{tab_2} indicate a consistent preference for the $\Lambda$CDM model over the $\xi$IDE scenario for all dataset combinations considered in this work. For the DESI DR1 dataset combinations, the logarithmic Bayes factors range from $\ln\mathcal{B}_{i,j}=4.18$ to $5.38$, corresponding to moderate-to-strong evidence in favor of the $\Lambda$CDM model according to the revised Jeffreys scale. In particular, the CMB + DESI DR1 + CC and CMB + DESI DR1 + CC + Pantheon$^{+}$ combinations yield $\ln\mathcal{B}_{i,j}=5.38$ and $5.35$, respectively, providing strong support for the $\Lambda$CDM scenario. For the DESI DR2 dataset combinations, the preference remains positive, with $\ln\mathcal{B}_{i,j}$ values ranging from $2.60$ to $4.16$, corresponding to moderate evidence in favor of the $\Lambda$CDM model. Although the strength of the preference decreases when DESI DR2 measurements are employed, the Bayesian evidence continues to favor the $\Lambda$CDM model over $\xi$IDE for all dataset combinations considered in this work.

In this work, we have investigated the phenomenological interacting dark energy model ($\xi$IDE) using combinations of compressed CMB likelihoods, DESI DR1 and DR2 BAO measurements, cosmic chronometers, and three independent SNe Ia compilations, namely Pantheon$^+$, DES-Dovekie, and Union3. Our analysis shows that the $\xi$IDE scenario provides a good description of the current cosmological observations and yields parameter constraints that remain largely consistent with the standard $\Lambda$CDM framework.

The interaction model generally favors slightly larger values of the Hubble parameter compared to $\Lambda$CDM, leading to a noticeable reduction in the tension with the local R22 measurement. However, the discrepancy is not completely removed for any of the dataset combinations considered in this work. The parameter $\xi$ is consistently constrained near the canonical value $\xi=3$, indicating that the current data do not provide strong evidence for a substantial departure from the standard cosmological evolution. Similarly, the dark energy equation-of-state parameter remains close to the cosmological constant value, with a mild preference for quintessence-like behavior.

The interaction parameter combination, $\xi+3w_{\rm X}$, is found to be negative for all dataset combinations, corresponding to an energy transfer from dark energy to dark matter. Nevertheless, the statistical significance of this preference remains low, with the strongest deviation from the non-interacting limit reaching only $1.52\sigma$. Consequently, the current observations provide only a weak indication of a non-zero dark-sector interaction.

Model comparison further supports this conclusion. Although the $\xi$IDE model yields a modest improvement in the fit relative to $\Lambda$CDM, the corresponding significance remains below the threshold required for moderate evidence. Moreover, the Bayesian evidence consistently favors the simpler $\Lambda$CDM model over the interacting scenario. Therefore, while the $\xi$IDE framework remains a viable extension of the standard cosmological model and offers a possible mechanism for reducing the Hubble tension, current cosmological observations do not provide compelling evidence for a dark-sector interaction.

%\section*{Acknowledgements}

%\bibliographystyle{elsarticle-num}
\bibliography{mybib.bib}

@article{SupernovaSearchTeam:1998fmf,
    author = "Riess, Adam G. and others",
    collaboration = "Supernova Search Team",
    title = "{Observational evidence from supernovae for an accelerating universe and a cosmological constant}",
    journal = "Astron. J.",
    volume = "116",
    pages = "1009--1038",
    year = "1998"
}

@article{SupernovaCosmologyProject:1998vns,
    author = "Perlmutter, S. and others",
    collaboration = "Supernova Cosmology Project",
    title = "{Measurements of $\Omega$ and $\Lambda$ from 42 High Redshift Supernovae}",
    journal = "Astrophys. J.",
    volume = "517",
    pages = "565--586",
    year = "1999"
}

@article{Padmanabhan:2002ji,
    author = "Padmanabhan, T.",
    title = "{Cosmological constant: The Weight of the vacuum}",
    journal = "Phys. Rept.",
    volume = "380",
    pages = "235--320",
    year = "2003"
}

@article{Copeland:2006wr,
    author = "Copeland, Edmund J. and Sami, M. and Tsujikawa, Shinji",
    title = "{Dynamics of dark energy}",
    journal = "Int. J. Mod. Phys. D",
    volume = "15",
    pages = "1753--1936",
    year = "2006"
}

@article{Zlatev:1998tr,
    author = "Zlatev, Ivaylo and Wang, Li-Min and Steinhardt, Paul J.",
    title = "{Quintessence, cosmic coincidence, and the cosmological constant}",
    journal = "Phys. Rev. Lett.",
    volume = "82",
    pages = "896--899",
    year = "1999"
}

@article{ratra1988cosmological,
  title={Cosmological consequences of a rolling homogeneous scalar field},
  author={Ratra, Bharat and Peebles, Philip JE},
  journal={Physical Review D},
  volume={37},
  number={12},
  pages={3406},
  year={1988},
  publisher={APS}
}

@article{caldwell2002phantom,
  title={A phantom menace? Cosmological consequences of a dark energy component with super-negative equation of state},
  author={Caldwell, Robert R},
  journal={Physics Letters B},
  volume={545},
  number={1-2},
  pages={23--29},
  year={2002},
  publisher={Elsevier}
}

@article{feng2006quintom,
  title={The quintom model of dark energy},
  author={Feng, Bo},
  journal={arXiv preprint astro-ph/0602156},
  year={2006}
}

@article{Armendariz-Picon:2000ulo,
    author = "Armendariz-Picon, C. and Mukhanov, Viatcheslav F. and Steinhardt, Paul J.",
    title = "{Essentials of k essence}",
    journal = "Phys. Rev. D",
    volume = "63",
    pages = "103510",
    year = "2001"
}

@article{Chiba:2002mw,
    author = "Chiba, Takeshi",
    title = "{Tracking K-essence}",
    journal = "Phys. Rev. D",
    volume = "66",
    pages = "063514",
    year = "2002"
}

@article{Weinberg:1988cp,
    author = "Weinberg, Steven",
    editor = "Hsu, Jong-Ping and Fine, D.",
    title = "{The Cosmological Constant Problem}",
    reportNumber = "UTTG-12-88",
    journal = "Rev. Mod. Phys.",
    volume = "61",
    pages = "1--23",
    year = "1989"
}

@article{Higson:2018cwj,
    author = "Higson, Edward and Handley, Will and Hobson, Michael and Lasenby, Anthony",
    title = "{Dynamic nested sampling: an improved algorithm for parameter estimation and evidence calculation}",
    journal = "Stat. Comput.",
    volume = "29",
    number = "5",
    pages = "891--913",
    year = "2018"
}

@article{Ashton:2022grj,
    author = "Ashton, Greg and others",
    title = "{Nested sampling for physical scientists}",
    journal = "Nature",
    volume = "2",
    year = "2022"
}

@article{speagle2020dynesty,
  title={dynesty: a dynamic nested sampling package for estimating Bayesian posteriors and evidences},
  author={Speagle, Joshua S},
  journal={Monthly Notices of the Royal Astronomical Society},
  volume={493},
  number={3},
  pages={3132--3158},
  year={2020},
  publisher={Oxford University Press}
}

@article{DESI:2024aqx,
    author = "Calderon, R. and others",
    collaboration = "DESI",
    title = "{DESI 2024: reconstructing dark energy using crossing statistics with DESI DR1 BAO data}",
    journal = "JCAP",
    volume = "10",
    pages = "048",
    year = "2024"
}

@article{DESI:2024kob,
    author = "Lodha, K. and others",
    collaboration = "DESI",
    title = "{DESI 2024: Constraints on physics-focused aspects of dark energy using DESI DR1 BAO data}",
    journal = "Phys. Rev. D",
    volume = "111",
    number = "2",
    pages = "023532",
    year = "2025"
}

@article{DESI:2025fii,
    author = "Lodha, K. and others",
    collaboration = "DESI",
    title = "{Extended dark energy analysis using DESI DR2 BAO measurements}",
    journal = "Phys. Rev. D",
    volume = "112",
    number = "8",
    pages = "083511",
    year = "2025"
}

@article{ChaudharyEPJc,
    author = "Chaudhary, Himanshu and Capozziello, Salvatore and Sharma, Vipin Kumar and G{\'o}mez-Vargas, Isidro and Mustafa, G.",
    title = "{Evidence for evolving dark energy from DESI DR2 BAO and Pantheon$^+$, DES-Dovekie, and Union3}",
    journal = "Eur. Phys. J. C",
    volume = "86",
    number = "5",
    pages = "564",
    year = "2026"
}

@article{ChaudharyPDU,
    author = "Capozziello, Salvatore and Chaudhary, Himanshu and Harko, Tiberiu and Mustafa, Ghulam",
    title = "{Is dark energy dynamical in the DESI era? A critical review}",
    journal = "Phys. Dark Univ.",
    volume = "51",
    pages = "102196",
    year = "2026"
}

@article{SharmaJHEAp,
    author = "Sharma, Vipin kumar and Chaudhary, Himanshu and Kolekar, Sanved",
    title = "{Probing generalized emergent dark energy with DESI DR2}",
    eprint = "2507.00835",
    archivePrefix = "arXiv",
    primaryClass = "astro-ph.CO",
    doi = "10.1016/j.jheap.2025.100518",
    journal = "JHEAp",
    volume = "51",
    pages = "100518",
    year = "2026"
}

@article{ChaudharyJHEAp,
    author = "Chaudhary, Himanshu and Capozziello, Salvatore and Praharaj, Subhrat and Pacif, Shibesh Kumar Jas and Mustafa, G.",
    title = "{Is the $\Lambda$CDM model in crisis?}",
    journal = "JHEAp",
    volume = "50",
    pages = "100507",
    year = "2026"
}

@article{ChaudharyAPJs,
    author = "Chaudhary, Himanshu and Sharma, Vipin Kumar and Capozziello, Salvatore and Mustafa, G.",
    title = "{Probing Departures from $\Lambda$CDM by Late-time Datasets}",
    journal = "Astrophys. J. Suppl.",
    volume = "283",
    number = "2",
    pages = "73",
    year = "2026"
}

@article{ChaudharyAaA,
    author = "Capozziello, Salvatore and Chaudhary, Himanshu and Mustafa, G. and Pacif, S. K. J.",
    title = "{Evidence of dynamical dark energy found via the DESI DR2 Lyman$\alpha$ forest}",
    journal = "Astron. Astrophys.",
    volume = "709",
    pages = "A258",
    year = "2026"
}

@article{ChaudharyAPj,
    author = "Chaudhary, Himanshu and Capozziello, Salvatore and Sharma, Vipin Kumar and Mustafa, Ghulam",
    title = "{Does DESI DR2 Challenge $\Lambda$CDM Paradigm?}",
    journal = "Astrophys. J.",
    volume = "992",
    number = "2",
    pages = "194",
    year = "2025"
}

@article{RitikaJHEAp,
    author = "Nagpal, Ritika and Chaudhary, Himanshu and Gupta, Harshita and Pacif, S. K. J.",
    title = "{Late-time constraints on dynamical dark energy models using DESI DR2, Type Ia supernova, and CC measurements}",
    doi = "10.1016/j.jheap.2025.100396",
    journal = "JHEAp",
    volume = "47",
    pages = "100396",
    year = "2025"
}

@article{lewis2025getdist,
  title={GetDist: a Python package for analysing Monte Carlo samples},
  author={Lewis, Antony},
  journal={Journal of Cosmology and Astroparticle Physics},
  volume={2025},
  number={08},
  pages={025},
  year={2025},
  publisher={IOP Publishing}
}

@article{Jimenez:2001gg,
    author = "Jimenez, Raul and Loeb, Abraham",
    title = "{Constraining cosmological parameters based on relative galaxy ages}",
    journal = "Astrophys. J.",
    volume = "573",
    pages = "37--42",
    year = "2002"
}

@article{moresco2012improved,
  title={Improved constraints on the expansion rate of the Universe up to z~ 1.1 from the spectroscopic evolution of cosmic chronometers},
  author={Moresco, Michele and Cimatti, Andrea and Jimenez, Raul and Pozzetti, Lucia and Zamorani, Gianni and Bolzonella, Micoll and Dunlop, James and Lamareille, Fabrice and Mignoli, Marco and Pearce, Henry and others},
  journal={Journal of Cosmology and Astroparticle Physics},
  volume={2012},
  number={08},
  pages={006--006},
  year={2012}
}

@article{Moresco:2015cya,
    author = "Moresco, Michele",
    title = "{Raising the bar: new constraints on the Hubble parameter with cosmic chronometers at z {\ensuremath{\sim}} 2}",
    journal = "Mon. Not. Roy. Astron. Soc.",
    volume = "450",
    number = "1",
    pages = "L16--L20",
    year = "2015"
}

@article{Moresco:2016mzx,
    author = "Moresco, Michele and Pozzetti, Lucia and Cimatti, Andrea and Jimenez, Raul and Maraston, Claudia and Verde, Licia and Thomas, Daniel and Citro, Annalisa and Tojeiro, Rita and Wilkinson, David",
    title = "{A 6{\%} measurement of the Hubble parameter at $z\sim0.45$: direct evidence of the epoch of cosmic re-acceleration}",
    journal = "JCAP",
    volume = "05",
    pages = "014",
    year = "2016"
}

@article{Moresco:2018xdr,
    author = "Moresco, Michele and Jimenez, Raul and Verde, Licia and Pozzetti, Lucia and Cimatti, Andrea and Citro, Annalisa",
    title = "{Setting the Stage for Cosmic Chronometers. I. Assessing the Impact of Young Stellar Populations on Hubble Parameter Measurements}",
    journal = "Astrophys. J.",
    volume = "868",
    number = "2",
    pages = "84",
    year = "2018"
}

@article{Moresco:2020fbm,
    author = "Moresco, Michele and Jimenez, Raul and Verde, Licia and Cimatti, Andrea and Pozzetti, Lucia",
    title = "{Setting the Stage for Cosmic Chronometers. II. Impact of Stellar Population Synthesis Models Systematics and Full Covariance Matrix}",
    journal = "Astrophys. J.",
    volume = "898",
    number = "1",
    pages = "82",
    year = "2020"
}

@article{adame2025desi,
  title={DESI 2024 VI: cosmological constraints from the measurements of baryon acoustic oscillations},
  author={Adame, AG and others},
  journal={Journal of Cosmology and Astroparticle Physics},
  volume={2025},
  number={02},
  pages={021},
  year={2025},
  publisher={IOP Publishing}
}

@article{benisty2024late,
  title={Late-time constraints on interacting dark energy: Analysis independent of $H_0$, $r_d$, and $M_B$},
  author={Benisty, David and Pan, Supriya and Staicova, Denitsa and Di Valentino, Eleonora and Nunes, Rafael C},
  journal={Astronomy \& Astrophysics},
  volume={688},
  pages={A156},
  year={2024},
  publisher={EDP sciences}
}

@article{Giare:2024smz,
    author = "Giar{\`e}, William and Sabogal, Miguel A. and Nunes, Rafael C. and Di Valentino, Eleonora",
    title = "{Interacting Dark Energy after DESI Baryon Acoustic Oscillation Measurements}",
    journal = "Phys. Rev. Lett.",
    volume = "133",
    number = "25",
    pages = "251003",
    year = "2024"
}

@article{Clemson:2011an,
    author = "Clemson, Timothy and Koyama, Kazuya and Zhao, Gong-Bo and Maartens, Roy and Valiviita, Jussi",
    title = "{Interacting Dark Energy -- constraints and degeneracies}",
    journal = "Phys. Rev. D",
    volume = "85",
    pages = "043007",
    year = "2012"
}

@article{Li:2014eha,
    author = "Li, Yun-He and Zhang, Jing-Fei and Zhang, Xin",
    title = "{Parametrized Post-Friedmann Framework for Interacting Dark Energy}",
    journal = "Phys. Rev. D",
    volume = "90",
    number = "6",
    pages = "063005",
    year = "2014"
}

@article{Xia:2016vnp,
    author = "Xia, Dong-Mei and Wang, Sai",
    title = "{Constraining interacting dark energy models with latest cosmological observations}",
    journal = "Mon. Not. Roy. Astron. Soc.",
    volume = "463",
    number = "1",
    pages = "952--956",
    year = "2016"
}

@article{van2025linear,
  title={I. Linear interacting dark energy: Analytical solutions and theoretical pathologies},
  author={Van Der Westhuizen, Marcel and Abebe, Amare and Di Valentino, Eleonora},
  journal={Physics of the Dark Universe},
  pages={102119},
  year={2025},
  publisher={Elsevier}
}

@article{vanderWesthuizen:2025mnw,
    author = "van der Westhuizen, Marcel and Abebe, Amare and Di Valentino, Eleonora",
    title = "{II. Non-linear interacting dark energy: Analytical solutions and theoretical pathologies}",
    journal = "Phys. Dark Univ.",
    volume = "50",
    pages = "102120",
    year = "2025"
}

@article{zhai2023consistent,
  title={A consistent view of interacting dark energy from multiple CMB probes},
  author={Zhai, Yuejia and Giar{\`e}, William and van de Bruck, Carsten and Di Valentino, Eleonora and Mena, Olga and Nunes, Rafael C},
  journal={Journal of Cosmology and Astroparticle Physics},
  volume={2023},
  number={07},
  pages={032},
  year={2023},
  publisher={IOP Publishing}
}

@article{escamilla2023model,
  title={Model-independent reconstruction of the interacting dark energy kernel: Binned and Gaussian process},
  author={Escamilla, Luis A and Akarsu, {\"O}zg{\"u}r and Di Valentino, Eleonora and Vazquez, J Alberto},
  journal={Journal of Cosmology and Astroparticle Physics},
  volume={2023},
  number={11},
  pages={051},
  year={2023},
  publisher={IOP Publishing}
}

@article{forconi2024double,
  title={A double take on early and interacting dark energy from JWST},
  author={Forconi, Matteo and Giar{\`e}, William and Mena, Olga and Di Valentino, Eleonora and Melchiorri, Alessandro and Nunes, Rafael C and others},
  journal={Journal of Cosmology and Astroparticle Physics},
  volume={2024},
  number={05},
  pages={097},
  year={2024},
  publisher={IOP Publishing}
}

@article{sabogal2024quantifying,
  title={Quantifying the $S_8$ tension and evidence for interacting dark energy from redshift-space distortion measurements},
  author={Sabogal, Miguel A and Silva, Emanuelly and Nunes, Rafael C and Kumar, Suresh and Di Valentino, Eleonora and Giar{\`e}, William},
  journal={Physical Review D},
  volume={110},
  number={12},
  pages={123508},
  year={2024},
  publisher={APS}
}

@article{giare2024tightening,
  title={Tightening the reins on nonminimal dark sector physics: Interacting dark energy with dynamical and nondynamical equation of state},
  author={Giar{\`e}, William and Zhai, Yuejia and Pan, Supriya and Di Valentino, Eleonora and Nunes, Rafael C and van de Bruck, Carsten},
  journal={Physical Review D},
  volume={110},
  number={6},
  pages={063527},
  year={2024},
  publisher={APS}
}

@article{silva2024non,
  title={Non-linear matter power spectrum modeling in interacting dark energy cosmologies},
  author={Silva, Emanuelly and Z{\'u}{\~n}iga-Bola{\~n}o, Ubaldo and Nunes, Rafael C and Di Valentino, Eleonora},
  journal={The European Physical Journal C},
  volume={84},
  number={10},
  pages={1104},
  year={2024},
  publisher={Springer}
}

@article{zhai2025low,
    author = "Zhai, Yuejia and de Cesare, Marco and van de Bruck, Carsten and Di Valentino, Eleonora and Wilson-Ewing, Edward",
    title = "{A low-redshift preference for an interacting dark energy model}",
    journal = "JCAP",
    volume = "11",
    pages = "010",
    year = "2025"
}

@article{silva2025new,
  title={New constraints on interacting dark energy from DESI DR2 BAO observations},
  author={Silva, Emanuelly and Sabogal, Miguel A and Scherer, Mateus and Nunes, Rafael C and Di Valentino, Eleonora and Kumar, Suresh},
  journal={Physical Review D},
  volume={111},
  number={12},
  pages={123511},
  year={2025},
  publisher={APS}
}

@article{KumarSharma2026,
  title={Does DESI DR2 hint at energy transfer from dark energy to dark matter?},
  author={Sharma, Vipin Kumar and Kumar, Ravinder and Chaudhary, Himanshu and Atamurotov, Farruh},
  journal={Journal of High Energy Astrophysics},
  volume={53},
  pages={100644},
  year={2026},
  publisher={Elsevier}
}

@article{vanderWesthuizen:2025rip,
    author = "van der Westhuizen, Marcel and Abebe, Amare and Di Valentino, Eleonora",
    title = "{III. Interacting Dark Energy: Summary of models, Pathologies, and Constraints}",
    journal = "Phys. Dark Univ.",
    volume = "50",
    pages = "102121",
    year = "2025"
}

@article{Li:2025owk,
    author = "Li, Tian-Nuo and Du, Guo-Hong and Li, Yun-He and Wu, Peng-Ju and Jin, Shang-Jie and Zhang, Jing-Fei and Zhang, Xin",
    title = "{Probing the sign-changeable interaction between dark energy and dark matter with DESI baryon acoustic oscillations and DES supernovae data}",
    journal = "Sci. China Phys. Mech. Astron.",
    volume = "69",
    number = "1",
    pages = "210413",
    year = "2026"
}

@article{DESI:2025zgx,
    author = "Abdul Karim, M. and others",
    collaboration = "DESI",
    title = "{DESI DR2 results. II. Measurements of baryon acoustic oscillations and cosmological constraints}",
    journal = "Phys. Rev. D",
    volume = "112",
    number = "8",
    pages = "083515",
    year = "2025"
}

@article{Amendola:1999er,
    author = "Amendola, Luca",
    title = "{Coupled quintessence}",
    journal = "Phys. Rev. D",
    volume = "62",
    pages = "043511",
    year = "2000"
}

@article{Zimdahl:2001ar,
    author = "Zimdahl, Winfried and Pavon, Diego",
    title = "{Interacting quintessence}",
    journal = "Phys. Lett. B",
    volume = "521",
    pages = "133--138",
    year = "2001"
}

@article{Chimento:2003iea,
    author = "Chimento, Luis P. and Jakubi, Alejandro S. and Pavon, Diego and Zimdahl, Winfried",
    title = "{Interacting quintessence solution to the coincidence problem}",
    journal = "Phys. Rev. D",
    volume = "67",
    pages = "083513",
    year = "2003"
}

@article{Guo:2004xx,
    author = "Guo, Zong-Kuan and Cai, Rong-Gen and Zhang, Yuan-Zhong",
    title = "{Cosmological evolution of interacting phantom energy with dark matter}",
    journal = "JCAP",
    volume = "05",
    pages = "002",
    year = "2005"
}

@article{cao2011testing,
  title={Testing the phenomenological interacting dark energy with observational H (z) data},
  author={Cao, Shuo and Liang, Nan and Zhu, Zong-Hong},
  journal={Monthly Notices of the Royal Astronomical Society},
  volume={416},
  number={2},
  pages={1099--1104},
  year={2011},
  publisher={The Royal Astronomical Society}
}

@article{Wei:2006ut,
    author = "Wei, Hao and Zhang, Shuang Nan",
    title = "{Observational H(z) Data and Cosmological Models}",
    journal = "Phys. Lett. B",
    volume = "644",
    pages = "7--15",
    year = "2007"
}

@article{brout2022pantheon+,
  title={The Pantheon+ analysis: cosmological constraints},
  author={Brout, Dillon and Scolnic, Dan and Popovic, Brodie and Riess, Adam G and Carr, Anthony and Zuntz, Joe and Kessler, Rick and Davis, Tamara M and Hinton, Samuel and Jones, David and others},
  journal={The Astrophysical Journal},
  volume={938},
  number={2},
  pages={110},
  year={2022},
  publisher={IOP Publishing}
}

@article{abbott2024dark,
  title={The dark energy survey: Cosmology results with\~{} 1500 new high-redshift Type Ia supernovae using the full 5-year dataset},
  author={Abbott, TMC and Acevedo, M and Aguena, M and Alarcon, A and Allam, S and Alves, O and Amon, A and Andrade-Oliveira, F and Annis, J and Armstrong, P and others},
  journal={arXiv preprint arXiv:2401.02929},
  year={2024}
}

@article{Rubin:2023jdq,
    author = "Rubin, David and others",
    title = "{Union Through UNITY: Cosmology with 2,000 SNe Using a Unified Bayesian Framework}",
    journal = "Astrophys. J.",
    volume = "986",
    number = "2",
    pages = "231",
    year = "2025"
}

@article{DES:2025sig,
    author = "Popovic, B. and others",
    collaboration = "DES",
    title = "{The Dark Energy Survey Supernova Program: A Reanalysis Of Cosmology Results And Evidence For Evolving Dark Energy With An Updated Type Ia Supernova Calibration}",
    eprint = "2511.07517",
    archivePrefix = "arXiv",
    primaryClass = "astro-ph.CO",
    reportNumber = "FERMILAB-PUB-25-0842-CSAID-PPD",
    doi = "10.1093/mnras/stag632",
    journal = "Mon. Not. Roy. Astron. Soc.",
    volume = "548",
    pages = "stag632",
    year = "2026"
}

@article{kass1995bayes,
  title={Bayes factors},
  author={Kass, Robert E and Raftery, Adrian E},
  journal={Journal of the american statistical association},
  volume={90},
  number={430},
  pages={773--795},
  year={1995},
  publisher={Taylor \& Francis}
}

@article{trotta2008bayes,
  title={Bayes in the sky: Bayesian inference and model selection in cosmology},
  author={Trotta, Roberto},
  journal={Contemporary Physics},
  volume={49},
  number={2},
  pages={71--104},
  year={2008},
  publisher={Taylor \& Francis}
}

@article{Dalal:2001dt,
    author = "Dalal, Neal and Abazajian, Kevork and Jenkins, Elizabeth Ellen and Manohar, Aneesh V.",
    title = "{Testing the cosmic coincidence problem and the nature of dark energy}",
    journal = "Phys. Rev. Lett.",
    volume = "87",
    pages = "141302",
    year = "2001"
}

@article{Pavon:2004xk,
    author = "Pavon, Diego and Sen, Somasri and Zimdahl, Winfried",
    title = "{CMB constraints on interacting cosmological models}",
    journal = "JCAP",
    volume = "05",
    pages = "009",
    year = "2004"
}

@article{Riess:2021jrx,
    author = "Riess, Adam G. and others",
    title = "{A Comprehensive Measurement of the Local Value of the Hubble Constant with $1~\mathrm{km\,s^{-1}\,Mpc^{-1}}$ Uncertainty from the Hubble Space Telescope and the SH0ES Team}",
    journal = "Astrophys. J. Lett.",
    volume = "934",
    number = "1",
    pages = "L7",
    year = "2022"
}

@article{Nong:2024bkr,
    author = "Nong, Xiao-Dong and Liang, Nan",
    title = "{Testing the Phenomenological Interacting Dark Energy Model with Gamma-Ray Bursts and Pantheon+ type Ia Supernovae}",
    journal = "Res. Astron. Astrophys.",
    volume = "24",
    number = "12",
    pages = "125003",
    year = "2024"
}

@article{Zhu:2025lrk,
    author = "Zhu, Ziyan and Jiang, Qingquan and Liu, Yu and Wu, Puxun and Liang, Nan",
    title = "{Cosmological constraints on the phenomenological interacting dark energy model with Fermi gamma-ray bursts and DESI DR2}",
    journal = "JHEAp",
    volume = "51",
    pages = "100534",
    year = "2026"
}

@article{Zheng:2022gfi,
    author = "Zheng, Jie and Chen, Yun and Xu, Tengpeng and Zhu, Zong-Hong",
    title = "{Investigating the dynamical models of cosmology with recent observations and upcoming gravitational-wave data}",
    journal = "Eur. Phys. J. Plus",
    volume = "137",
    number = "4",
    pages = "509",
    year = "2022"
}

@article{Wang:2016lxa,
    author = "Wang, B. and Abdalla, E. and Atrio-Barandela, F. and Pavon, D.",
    title = "{Dark Matter and Dark Energy Interactions: Theoretical Challenges, Cosmological Implications and Observational Signatures}",
    journal = "Rept. Prog. Phys.",
    volume = "79",
    number = "9",
    pages = "096901",
    year = "2016"
}

@article{Cao:2010fb,
    author = "Cao, Shuo and Liang, Nan and Zhu, Zong-Hong",
    title = "{Testing the phenomenological interacting dark energy with observational $H(z)$ data}",
    journal = "Mon. Not. Roy. Astron. Soc.",
    volume = "416",
    pages = "1099--1104",
    year = "2011"
}

@article{Salvatelli:2014zta,
    author = "Salvatelli, Valentina and Said, Najla and Bruni, Marco and Melchiorri, Alessandro and Wands, David",
    title = "{Indications of a late-time interaction in the dark sector}",
    journal = "Phys. Rev. Lett.",
    volume = "113",
    number = "18",
    pages = "181301",
    year = "2014"
}

@article{Chen:2010dk,
    author = "Chen, Yun and Zhu, Zong-Hong and Alcaniz, J. S. and Gong, Yungui",
    title = "{Using A Phenomenological Model to Test the Coincidence Problem of Dark Energy}",
    journal = "Astrophys. J.",
    volume = "711",
    pages = "439--444",
    year = "2010"
}

@article{Guo:2007zk,
    author = "Guo, Zong-Kuan and Ohta, Nobuyoshi and Tsujikawa, Shinji",
    title = "{Probing the Coupling between Dark Components of the Universe}",
    journal = "Phys. Rev. D",
    volume = "76",
    pages = "023508",
    year = "2007"
}

@article{Lebedev:2021tas,
    author = "Lebedev, Oleg and Smirnov, Fedor and Solomko, Timofey and Yoon, Jong-Hyun",
    title = "{Dark matter production and reheating via direct inflaton couplings: collective effects}",
    journal = "JCAP",
    volume = "10",
    pages = "032",
    year = "2021"
}

\end{document}